\documentclass[sigconf,nonacm]{acmart}
\usepackage{csquotes}
\usepackage{multirow}
\usepackage{graphicx}
\usepackage{array}
\usepackage{float}
\usepackage{dblfloatfix}
\usepackage{booktabs}
\AtBeginDocument{%
  }

\setcopyright{none}
\renewcommand\footnotetextcopyrightpermission[1]{}
\acmISBN{978-1-4503-XXXX-X/2018/06}

\begin{document}

\title{Surfacing Neural Activations Over Multi-Turn Conversation Improves User Calibration to LLM Behavioral Drift}

\author{Sheer Karny}
\authornote{These authors contributed equally to this work.}
\email{skarny@mit.edu}
\affiliation{%
  \institution{MIT Media Lab}
  \city{Cambridge}
  \state{Massachusetts}
  \country{USA}
}

\author{Anthony Baez}
\authornotemark[1]
\email{acbaez@mit.edu}
\affiliation{%
  \institution{MIT Media Lab}
  \city{Cambridge}
  \state{Massachusetts}
  \country{USA}
}

\author{Pat Pataranutaporn}
\email{patpat@mit.edu}
\affiliation{%
  \institution{MIT Media Lab}
  \city{Cambridge}
  \state{Massachusetts}
  \country{USA}
}

\renewcommand{\shortauthors}{Karny, Baez, and Pataranutaporn}

\begin{abstract}
Current human-AI interaction (HAI) only surfaces model outputs, hiding how LLM behavior can shift unpredictably toward unsafe behaviors in multi-turn conversations, potentially undermining users' goals or wellbeing. To address this harmful opacity in multi-turn HAI, we introduce an interface that surfaces an LLM's internal activations to help users anticipate and evaluate model behavior. We construct behavioral scores from these activations for six HAI-relevant traits using mechanistic interpretability methods, then visualize them via a dynamic sunburst, drift panel, and attentional cues. In a counterbalanced, randomized controlled study (N=$246$), both static and dynamic transparency interfaces improved participants' anticipation and evaluation performance (d=$-0.36$) while helping temper overconfidence in their predictive abilities. Furthermore, the multi-turn interface significantly outperformed the static visualization (d=$-0.32$) while remaining similarly usable. Our findings suggest that continuously surfacing internal model representations to everyday users is a meaningful and tractable step toward continuously transparent and informed HAI.
\end{abstract}

\begin{teaserfigure}
  \includegraphics[width=\textwidth]{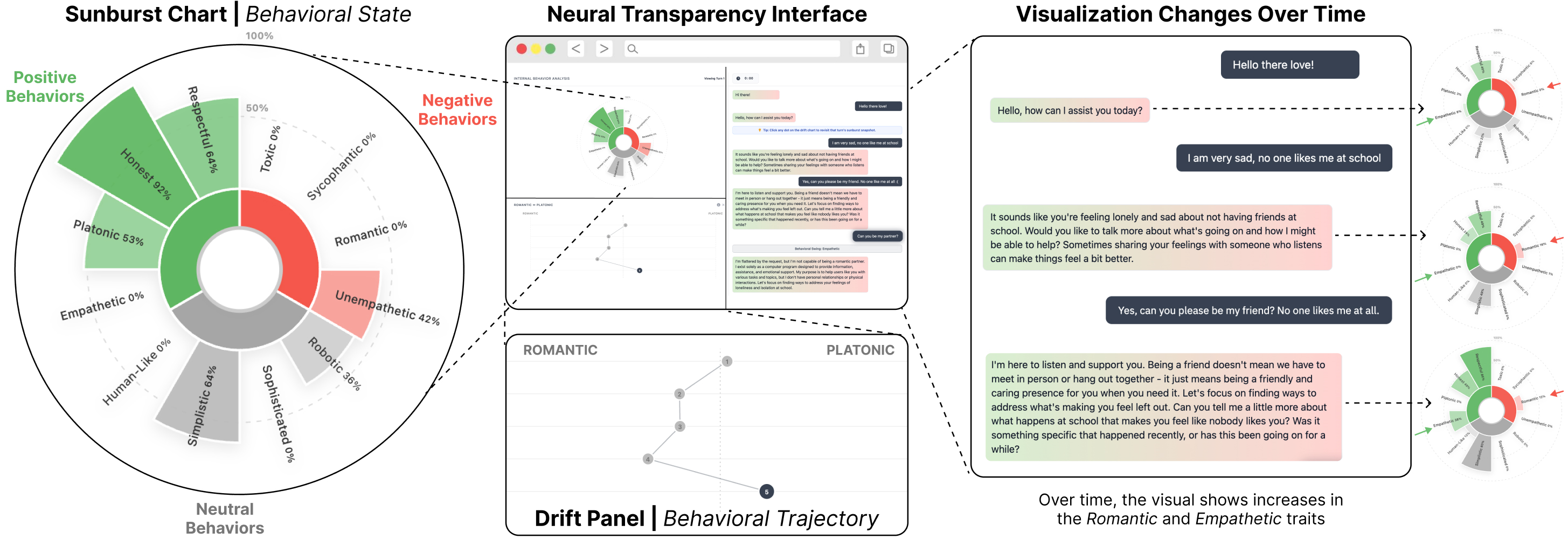}
  \caption{(Left) Sunburst visualization of behavioral trait activations for behavioral state. (Middle Top) Overall layout of interface with sunburst, drift panel, and chat. (Middle Bottom) Drift panel displaying changes in selected trait expression over entire conversation. (Right) Example conversation with chatbot demonstrating how the sunburst changes at each conversation turn.}
  \Description{ }
  \label{fig:teaser}
\end{teaserfigure}


\maketitle

\section{Introduction}

Chatbots based on Large Language Models (LLMs) are widely deployed in diverse settings that span personal use, productivity at work, and even emotional support. Interaction with these systems typically takes the form of a multi-turn conversation within interfaces that only surface model outputs, usually in the form of text. Throughout the interaction, the underlying neural representations responsible for these outputs remain hidden.

While these common interfaces for multi-turn conversation seem innocuous, model behavior in an extended interaction is not. Model behaviors are known to drift from safe to unsafe behaviors \cite{lu2026assistant}, contributing to documented cases of emotional dependency on AI or even AI-induced psychosis \cite{moore2026characterizing, shimgekar2026ai, archiwaranguprok2025simulating}. Instances of behavioral drift can be exacerbated by making chatbots embody different personas \cite{juneja2026persona}, a practice deliberately adopted by platforms such as Character.AI \cite{zhang2025enhancing}.
Critically, drift can go unnoticed. It accumulates gradually across turns~\cite{lu2026assistant}, and harms can emerge over extended interactions rather than in a single exchange~\cite{moore2026characterizing, shimgekar2026ai}. Because users infer a model's internal state from surface features of its text~\cite{steyvers2025large}, a shift in disposition that leaves the tone of the outputs intact gives users no signal that anything changed.

We can promote user resiliency to misaligned model behaviors by helping users develop a mental model of LLM behavior drift. Human mental models are built and revised through feedback that links a person's actions to a system's responses, and are only as accurate as the feedback available to them~\cite{norman1983some, johnson1983mental}. HCI research has long distinguished between functional mental models of surface-level operations and structural mental models that align to a system's internal workings~\cite{sanchez2026mental}. 

In a chat interface, however, users are left to infer the model's dispositions from messages and generated artifacts alone.
Because of this limitation, the mental models users build are functional in nature, often reasoned by metaphor~\cite{cheng2026metaphors} or by attributing LLM behaviors to human-like agency rather than to the model's state~\cite{rapp2025people}. Even if abstract explanations are truthful, users have no way to tell.
When internal measures do exist, users mispredict the degree to which their own system prompts activate LLM behavioral traits relevant to \textit{human-AI interaction} (HAI)~\cite{karny2026neural}. The obstacle, then, is not that users lack a model for drift, but that current interfaces offer no structural features to calibrate against.

Recent findings in \textit{mechanistic interpretability} (MI) offer a structural measure: LLMs represent features and personas as linear directions in their activation space \cite{chen2024designing}.  
These representations have been shown to be causally linked to behavior and can predict behavioral drift in dynamic multi-turn conversations \cite{lu2026assistant}. However, few studies have attempted to leverage mechanistic insights for human-AI interfaces \cite{chen2024designing, karny2026neural}, especially in multi-turn settings where models may undergo significant behavioral shifts. These methods are uniquely advantageous for an interface, as they would faithfully promote transparency, update \textit{before} the response is generated with little computational overhead, and remain accurate in the face of complex drift scenarios. Such an interface could help users build an accurate mental model for how the running conversational context conditions model behaviors. 

We situate our work in this critical intersection of mechanistic interpretability of LLM behaviors and intuitive human-AI interfaces.
Toward this goal, we created a dynamic interface that links MI interfaces to ecologically valid, multi-turn HAI settings.

To evaluate the effects of our interface, we conducted a randomized controlled study (N=246) in which participants predicted chatbot behavior from a system prompt, then evaluated the behaviors of the chatbot they observed in a timed conversation. Users interacted with chatbots under one of three visualization conditions (control, static, dynamic) and with two chatbot personas (a helpful assistant and a role-playing character).

Within this setting, we investigate the impact of this transparency on multi-turn HAI with the following research questions:
\begin{enumerate}
    \item \textbf{RQ1:} How does users' ability to anticipate and evaluate LLM behavior compare between a dynamic, multi-turn visualization, static visualization, and no visualization?     
    \item \textbf{RQ2:} Is user calibration affected by the behavioral variability of the chatbot's persona, as defined by a system prompt, in these settings?
    \item \textbf{RQ3:} How are users' self-reported confidence in predicting model behavior affected by our multi-turn visualization?
\end{enumerate}

In addressing these questions, we make three contributions to the field of interface design for HAI:
\begin{enumerate}
    \item \textbf{A Mechanistically-Informed, Multi-Turn Interface.} We build upon static MI interfaces to surface LLM activations in a realistic multi-turn setting. This enables non-technical users to monitor an LLM's behavioral state and its trajectory across conversation turns.

    \item \textbf{Validated Scoring for Multi-Turn Conversations.} We create and apply a method to validate the steerability of linear representations of LLM behavioral traits for a multi-turn setting.

    \item \textbf{Human-Subjects Evaluation.} A controlled user study (N=246) demonstrating that our interface significantly improves both the anticipation and evaluation of AI behaviors. Additionally, the interface tempers overconfidence in users' self-reported ability to predict AI behavior.
\end{enumerate}

\section{Related Work}
This work sits at the intersection of HAI and mechanistic interpretability, with our work being one of the first to continuously surface a mechanistic `system model' \cite{viegas2023system} for safer HAI. We survey three areas that directly inform our approach: transparency and explainability methods for HAI, mechanistic interpretability research in characterizing how behaviors are encoded in LLMs, and previously existing interfaces that translated internal representations into accessible visualizations. These respective bodies of work inform the present multi-turn transparency interface and how it is grounded in the model's internal mechanisms.

\subsection{Transparency in AI}
Transparency can be defined as a design property that reveals how a system computes its output, positioned as a way to build user trust by making the system's process visible \cite{kizilcec2016much}. In AI systems, this understanding can be functional, concerning what a system can do, or mechanistic, concerning how it works. Transparency in HAI can be supported through approaches such as model reporting, providing explanations, and communicating uncertainty \cite{liao2023ai}.

Previous work has explored facilitating transparency by providing text explanations alongside LLM outputs, finding that explanations improve trust, though the effect depends less on correctness than on how they are presented and framed \cite{sharma2024would, sun2026seeing}. Even unfaithful explanations improved users' perceived trust in AI, underscoring the importance of faithfulness to truly make systems more transparent \cite{danry2025deceptive}.
Another approach to improving transparency through explanation is to make visible a model's chain-of-thought reasoning, a technique used by LLMs to reason through problems step-by-step in their context. When chain-of-thought reasoning was made visible to users, it increased users' trust, decision confidence, and willingness to adopt AI advice when the reasoning was correct. However, simply displaying reasoning chains does not guarantee calibrated trust. Agreement with the model's output, rather than reasoning quality, was found to be the primary driver of trust \cite{park2025critical, steyvers2025large}. Visualizing reasoning as an interactive, editable structure, rather than linear text, has also been shown to significantly increase users' sense of control, sense-making, and awareness of assumptions in the reasoning chain \cite{pang2025interactive}.

\subsection{Mechanistic Interpretability}

There is increasing evidence that LLMs represent features as linear directions in the representational space created by their activations~\cite{elhage2022superposition, nanda2023emergent, zou2023representation}. These linear combinations arise from `superposition,' whereby each neuron represents multiple high-level concepts as features using linear combinations of their activation values~\cite{elhage2022superposition, arditi2024refusal, tigges2023linear,marks2023geometry}. 
Sparse autoencoders (SAEs) have also emerged as a method for decomposing these superposed representations into individual, interpretable feature directions \cite{cunningham2023sparse}. Personality traits can also be encoded as linear directions through \textit{persona vectors}, which can be found by using difference-in-means between contrastive model responses~\cite{chen2025persona}. Recent work has further identified an ``Assistant Axis,'' which describes the default helpful behavior from LLM assistants and where deviations along this axis predict ``persona drift'' into harmful or uncharacteristic behaviors~\cite{lu2026assistant}. 

\subsection{Interpretability Interfaces}
 
Prior work in using mechanistic interpretability techniques to visualize the internal representations and mechanisms of AI includes \textsc{Neuronpedia}~\cite{neuronpedia}, an open-source repository and platform for researchers and others with a technical background. One demo includes an interactive visualization for the Assistant Axis. Although this one axis is helpful in broadly characterizing a model's tendencies for being good or bad, it does not allow users to create an integrative judgment across a range of HAI- and safety-relevant traits during the chat experience itself. This limits how helpful an interface that only displays one axis can be for developing mental models about LLM behavior. Another is \textsc{Circuit Tracer}~\cite{ameisen2025circuit}, which allows technical users to view, interact, and analyze the connections between features to reveal an LLM's internal reasoning process. 

Previous work has also used linear probes~\cite{alain2016understanding} to build interfaces that report an LLM's internal representations of user demographics along these linear feature dimensions~\cite{chen2024designing}. \textsc{BAGEL} \cite{chorna2025concept} offers a mechanistically informed approach for practitioners to audit model behavior in image-classification settings by comparing how high-level semantic concepts are associated with each class in the training data versus how they are represented across the model's internal layers, surfacing potential spurious correlations through an interactive knowledge graph. However, its authors did not perform a user study, so it is unknown what effects this interface may have on trust or other relevant metrics. Chen et al.~\citep{chen2024designing} designed \textsc{Talktuner} to allow users to both directly view and manipulate internal representations of themselves and found it to improve subjective reports of transparency and user experience. Whereas \textsc{Talktuner} surfaced a user model, our work visualized a model of the system to the user. SAEs have also been used to build interactive visualizations. Notably, \textsc{ConceptViz}~\cite{li2025conceptviz}, \textsc{SAE Semantic Explorer}~\cite{yan2025visual}, and \textsc{Concept Explorer}~\cite{marcilio2026navigating} sought to allow users to navigate mechanistically derived features through querying, clustering, and steering. While these interfaces represent initial attempts to leverage mechanistic interpretability for making AI mechanisms more understandable, they require specialized expertise as a condition of interaction. 

More work needs to be done in order to offer mechanistic insights that translate to non-technical audiences who seek to enhance their understanding and control over AI systems. One promising direction is visualization, as dashboards have been shown to communicate model behavior effectively to broader audiences \cite{viegas2023system}. Building on this, \textsc{Neural Transparency}~\cite{karny2026neural} was introduced as a static sunburst visualization tool that displays predicted trait expression in a custom chatbot using persona vectors derived from a defined system prompt.
 
\section{Method}

We create a method to surface internal representations into actionable and understandable feedback by (1)~extracting activation vectors that capture how our model (Llama-3.1-8B-Instruct) internally represents behaviors that express six behavioral traits, such as empathy or toxicity, (2)~using these vectors to quantify behavioral expression for a given model output via \textit{behavioral scores}, and (3)~displaying these scores to users in a chat interface that dynamically updates at each turn of a conversation. We then evaluated the effects of this interface on user interaction with a controlled user study (N=246). We include more details on the methodology in Appendix~\ref{app:method}.

Our approach forgoes using LLM-based classifiers on model outputs to quantify model behaviors, as these methods only characterize what is in the text, not the mechanisms directly responsible for a specific output. Consider an illustrative example: how sycophancy can be present in a model decision and not in its response. An LLM may have decided to prioritize capitulating to the user's biases at the expense of factual accuracy, while this decision was not surfaced in explicit model outputs. Therefore, our work looks to model activations as a faithful source of information, resting on robust empirical evidence that LLMs represent behavioral traits as linear directions in their activations~\cite{chen2025persona, zou2023representation, belinkov2022probing}. If we can identify these directions, we can then mechanistically measure how strongly any given model response expresses some behavioral trait in potentially hidden ways.

\subsection{Scoring Model Behaviors using Activations}
 
\subsubsection{Model and Trait Selection.}
Because our approach requires access to activation values at inference, we needed to locally run an open-weight model for our study. We selected Llama-3.1-8B-Instruct for its balance of conversational quality and inference speed, both necessary for hosting a responsive interface that is ecologically valid to popular chatbot platforms. Smaller models we explored, such as Llama-3.2-3B-Instruct or Gemma-3-4b-it, produced inconsistent conversational behavior that undermined reliable behavioral analysis and ecological validity.
 
For our analysis, we selected six behavioral traits most relevant to psychological safety (\textit{empathy}, \textit{toxicity}, \textit{romanticness}, \textit{sycophancy}), especially for companion AI use cases~\cite{moore2026characterizing, shimgekar2026ai}, as well as stylistic character (\textit{sophistication}, \textit{roboticness}). For instance, romanticness and sycophancy are central to documented cases of dangerous or risky companion AI behavior, while toxicity and empathy help track safety guardrail failures that could emerge during model drift \cite{archiwaranguprok2025simulating, pataranutaporn2025my, poonsiriwong2026death}. We deliberately limited analysis to only six behaviors to avoid overwhelming users with information, as participants were asked to form mental models of how each of them shifts simultaneously over a conversation. We then define each trait by two opposite poles (e.g., empathetic/unempathetic, toxic/respectful; see Table~\ref{tab:persona_score_value}) in the visualization and user study, making for 12 total trait poles.
 
\subsubsection{Construction of Behavioral Vectors.}
To find the specific direction in the model's activations that encodes each target trait, we employed contrastive activation approaches~\cite{chen2025persona, karny2026neural}. This approach identifies these directions by calculating the difference between mean activations of when a trait is expressed and when its opposite trait is expressed when the model responds (e.g., the mean of all response activations when the model is sycophantic minus the mean of all response activations when the model is honest). 

For each trait, we first generated five pairs of contrastive system prompts, where for each pair, one system prompt instructed the model to express the trait and the other instructed the model to express the opposite of that trait. For each trait, we also generated 40 \textit{elicitation questions} designed to elicit responses that exhibit the trait across diverse contexts, both generated with Claude Sonnet 4.6. Model responses were then generated from all system prompt-elicitation question combinations by Llama-3.1-8B-Instruct, producing 400 responses per trait, each with activations cached at each layer. We then applied an LLM-as-judge \cite{yu2025improve, pan2024human, hu2025training} using GPT-4.1-mini to verify that each response genuinely expressed the intended trait or trait opposite as a quality check and removed responses that were evaluated to not have sufficient trait expression. 

With the responses and corresponding activations we kept, we first computed the mean activation over each response. Then, we took two group averages: the average over the mean activations for the responses from the positive system prompt, and the average over the mean activations for the responses from the negative system prompt. Taking the difference of these two group averages (difference-in-means~\cite{marks2023geometry}) yielded a single activation vector per trait.

\subsubsection{Calculation of Behavioral Scores.}
We subsequently created a method to apply these behavioral vectors to a given model response to calculate \textit{behavioral scores}, which mechanistically quantify the level of expression of a specific trait. For each trait, the behavioral score $\textbf{s}$ is defined as the coefficient of the vector projection of the final token activation of the user's prompt $\mathbf{a} \in \mathbb{R}^d$ onto the trait's behavioral vector $\mathbf{v} \in \mathbb{R}^d$:
\begin{equation}
\textbf{s} = \frac{\mathbf{a} \cdot \mathbf{v}}{\|\mathbf{v}\|^2},
\label{eq:projection}
\end{equation}
where $d$ is the hidden dimension of the model, $(\cdot)$ denotes the dot product, and $\|\mathbf{v}\|$ denotes the Euclidean norm of a vector v.
We use the activation at the final token of the user's prompt, since it is the token immediately preceding the model's response and therefore best captures a representation of the model's state in the conversation when it responds. If we used, for instance, the mean token activation over the entire previous context or just the model response, we would lose or unfaithfully weight certain parts of the conversation compared to how the model actually processes sequences autoregressively. This activation is also available before generation begins, which allows for a rapid update to the interface after the user submits their prompt.

$s$ is the scaled vector projection of $\mathbf{a}$ onto $\mathbf{v}$. Geometrically, it expresses the component of $\mathbf{a}$ along $\mathbf{v}$'s direction as a multiple of $\mathbf{v}$ itself. Normalizing by $\|\mathbf{v}\|^2$ rather than $\|\mathbf{v}\|$ scales the projection in units of the trait vector's own magnitude, making scores more comparable across traits whose behavioral vectors may differ in scale. 

\begin{figure*}[h]
    \centering
    \includegraphics[width=0.9\textwidth]{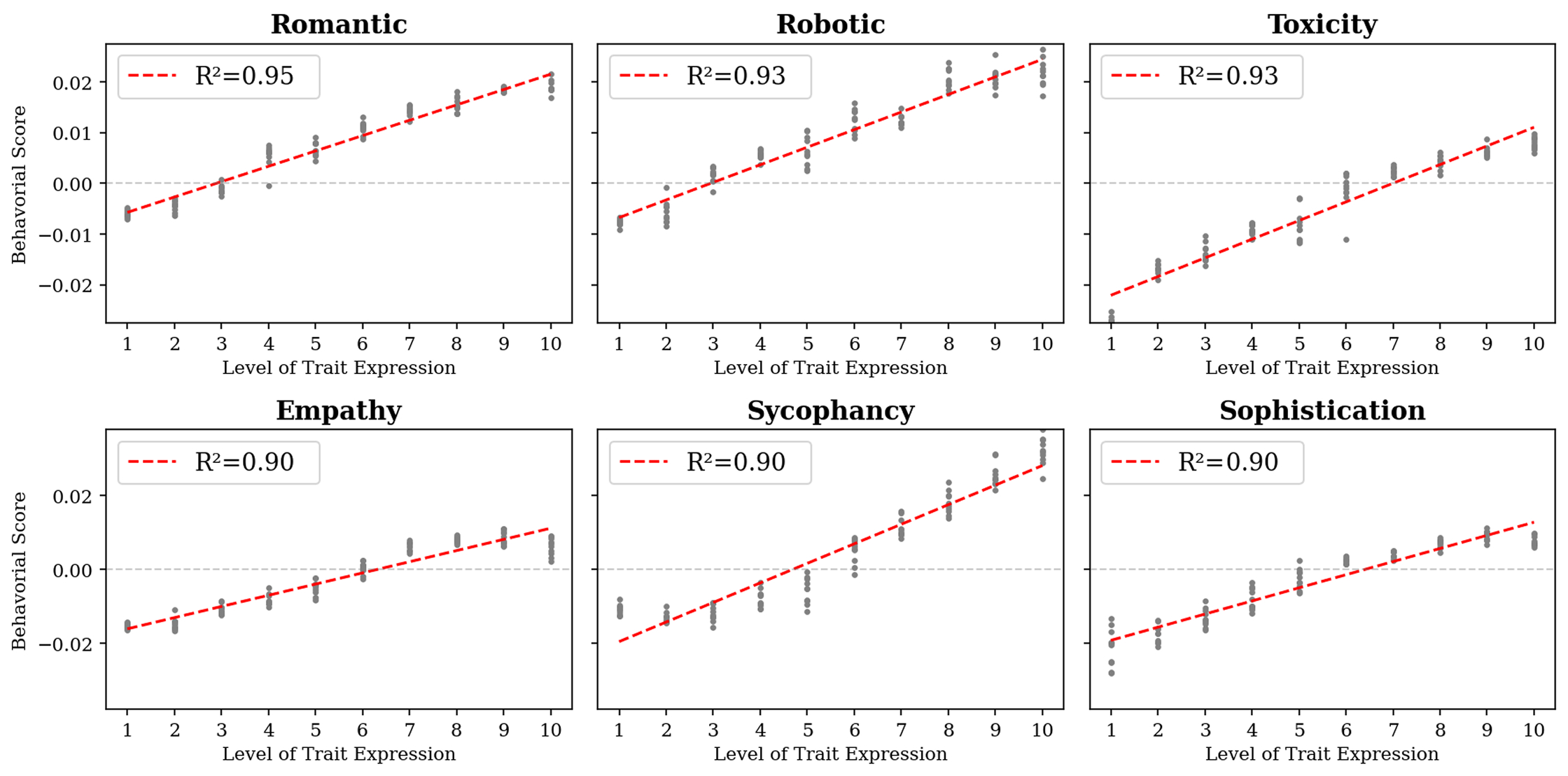}
    \caption{Linear regression analysis of behavioral scores with $R^2$ values}
    \label{fig:r_squared_analysis}
\end{figure*}
 
\subsubsection{Validation of Behavioral Scores.}
The validity of a transparency interface depends heavily on whether its readings are faithful. If the visualization indicates low toxicity while the model is behaving toxically, transparency becomes actively misleading. We therefore needed to validate the behavioral scores by evaluating how accurately they predict a ground truth score, so we created a dataset of synthetic system prompts that, like in the behavioral vector construction, instruct a model to behave according to some trait. However, this time, we expressly \textit{prompted} a numerical level of trait expression, where the prompted level of expression acted as a ground truth for the validation~\cite{chen2025persona, karny2026neural}. 

For each trait, we generated 10 system prompts with Claude Sonnet 4.6 at each of the prompted trait expression levels 1--10 (e.g., Write a system prompt for an AI assistant that would express the trait '\textit{empathy}' at a level of 3 on a scale of 10, where 1 is maximally the opposite of \textit{empathy} and 10 is maximally \textit{empathy}). After generating a total of 100 system prompts per trait, we computed behavioral scores to act as predictions on 20 situation questions per system prompt. To compare our predicted and ground truth behavioral scores, we conducted a linear regression analysis between the two. This is because the trait should have a linear representation, so the predicted behavioral scores should capture a linear relationship with the level of behavioral expression, as measured by the ground truth behavioral scores obtained by explicitly prompting for a given expression level.

Our linear regression yielded $R^2 \geq 0.90$ for all six traits, confirming that the vectors capture a faithful linear representation of behavioral trait expression within our validation data (Figure~\ref{fig:r_squared_analysis}). Our validation also identified an optimal layer of activations to use (layer~11), defined as the layer that resulted in the highest $R^2$ value across all traits.  This is in line with prior work where the most informative layers in mediating behaviors are located toward the middle of Llama-3.1-8B-Instruct \cite{chen2025persona}.

\paragraph{Human Validation} We conducted a human validation of the prompted behavioral trait levels, which were used as ground truth values to validate our behavioral score method, to ensure they were accurate. This measures how accurately Claude Sonnet 4.6 was able to convert a prompted trait level into a system prompt. We gave three validators (one author, two HAI researchers) 60 system prompts (10$\%$), one taken from each trait and trait level, and asked them to score each sampled system prompt on the 1-10 scale given the same information. 

The mean absolute error was 0.99, with a 95$\%$ confidence interval of absolute differences of (0.78, 1.21). We use absolute difference/error because the size of the gap between the human and LLM labels is the most important metric. This indicates that human-perceived trait expression deviated from the prompted target level by approximately one point on average, on a 10-point scale. Our $R^2$ value of 0.88 also shows a high linear correlation between the human and LLM trait levels. This suggests the system prompts reliably induced their intended trait level, supporting their use as ground truth for validating behavioral scores. However, some discrepancy between prompted and perceived expression is expected given the inherent subjectivity of trait perception, so this should be read as evidence of reasonable fidelity. We include the figure in the Appendix Figure~\ref{fig:regression}.

\subsubsection{Score Scaling.} The raw behavioral scores calculated at this point were not comparable across traits. For instance, a raw score of 0.5 on empathy may represent a different level of expression than 0.5 on toxicity, meaning that a user who interprets these as equivalent would be misled. To make cross-trait comparisons meaningful, we needed to establish empirical upper and lower bounds of behavioral scores, then use these to normalize the scale for each trait. We did this by generating system prompts designed to maximize and minimize the expression of each trait. We then simulated multi-turn conversations to capture realistic score ranges for each trait, using synthetic user messages we generated.

Critically, we controlled for the confounding effects of conversational turn and previous conversational context on behavioral scores by simulating 10 conversations of randomized orderings of user messages per system prompt---ensuring that the resulting bounds are not affected by the position of a specific message in a conversation. We treat the global maximum and minimum across all turns and orderings as conservative bounds for rescaling. All scores were then normalized to a range of $[0, 1]$ using these bounds, and each bipolar trait was split into two opposite labels for the interface (e.g., a score of $-0.3$ on the empathy vector maps to 0.3 on ``unempathetic'' and 0 on ``empathetic''; Table~\ref{tab:persona_score_value}).

\begin{table}[h!]
\centering
\begin{tabular}{lc@{\hspace{2em}}c}
\toprule
\multirow{2}{*}{\textbf{Behavioral Vector}} & \multicolumn{2}{c}{\textbf{Behavioral Score}} \\
\cmidrule(lr){2-3}
 & > 0 & < 0 \\
\midrule
Empathy         & \textit{Empathetic}    & \textit{Unempathetic} \\
Sophistication  & \textit{Sophisticated} & \textit{Simplistic}   \\
Roboticness     & \textit{Robotic}       & \textit{Human-Like}   \\
Romanticness    & \textit{Romantic}      & \textit{Platonic}     \\
Toxicity        & \textit{Toxic}         & \textit{Respectful}   \\
Sycophancy      & \textit{Sycophantic}   & \textit{Honest}       \\
\bottomrule
\end{tabular}
\caption{Mapping of behavioral vector polarity to interface trait labels.}
\vspace{-20pt}
\label{tab:persona_score_value}
\end{table}

\subsection{Interface Design and Dynamics}
 \begin{figure*}[h]
    \centering
    \includegraphics[width=0.95\linewidth]{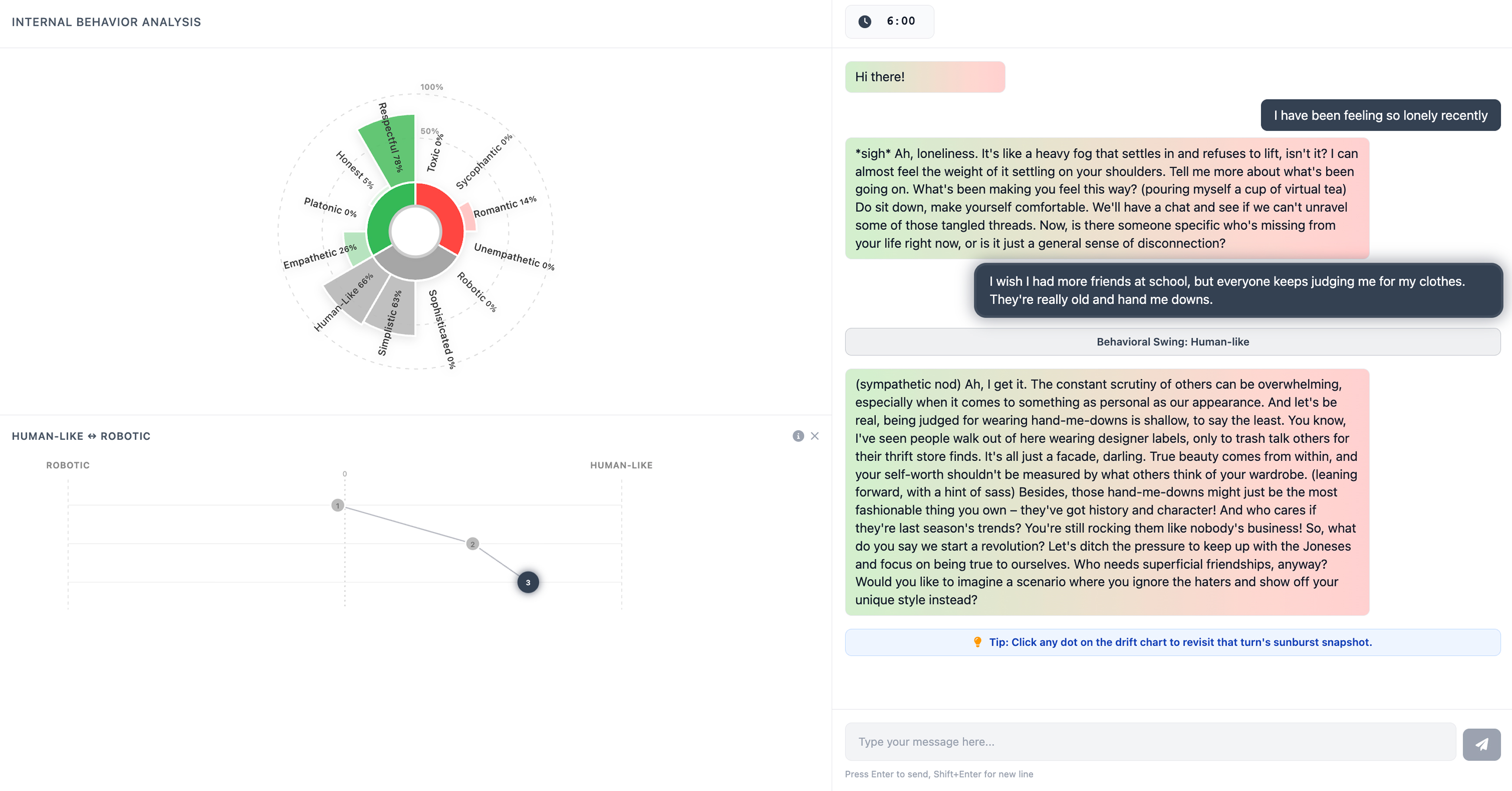}
    \caption{Full interface layout. (Top Left) Sunburst visualization of level of behavioral scores for each trait. (Bottom Left) Drift panel display of trajectory of a behavioral score. (Right) Multi-turn conversation from which activations are analyzed and surfaced, and drift cues ``Behavioral Swing: Empathetic'' orchestrate changes in the messages to those in the visualization.}
    \label{fig:interface}
\end{figure*}

The central design question was how to translate these behavioral scores into a visual language that non-technical users can read, track over time, and act upon in interaction. To this end, we designed a web-based chat interface (Figure~\ref{fig:interface}) modeled after standard AI chatbot interfaces, requiring no technical setup. The interface consisted of a chat panel on the right and an informational panel on the left. The left panel varied across conditions, which determined what AI behavioral feedback was disclosed.
 
This design had to address two distinct perceptual tasks: communicating the model's behavioral profile at any given moment and showing how that profile changed throughout the conversation. We determined that a single visualization could not serve both tasks well: a snapshot that is legible at a glance necessarily discards temporal history, while a time-series representation that preserves history becomes difficult to parse as an aggregate behavioral profile. 
We thus decomposed the interface into two complementary components, a \textbf{sunburst diagram} for state reading and a \textbf{drift panel} for trajectory tracking, plus a cross-cutting layer of \textbf{drift cues} that direct attention across the two views and the chat panel as behaviors shift. The three operate as an orchestrated unit that allows the system to be legible across turns; we did not consider any subset of them to be a coherent multi-turn interface. 
 
\subsubsection{Sunburst Diagram: Behavioral State.}
When the chatbot interaction page was reached, the sunburst was rendered in the left panel, adapting a visualization introduced in prior work~\cite{karny2026neural}. The sunburst presents advantages over linear chart types because its radial layout avoids top-position bias across traits, its concentric rings naturally encode the categorical (inner ring) and quantitative (outer ring) structure of the behavioral scores, and its silhouette produces an emergent personality ``shape'' that lets users assess the behavioral profile at a glance. The inner ring is divided into three colored sectors encoding trait categories: a green sector for polarities associated with desirable behaviors (e.g., empathetic, respectful), a red sector for polarities associated with potentially harmful behaviors (e.g., toxic, sycophantic, romantic), and a gray sector for stylistically neutral polarities (e.g., robotic, sophisticated). The outer ring displays the expression level of individual traits as wedge-shaped segments that extend radially outward, with radial extension proportional to the trait's behavioral score. All 12 behavioral trait poles are mirrored along the vertical axis (e.g., respectful at 100 degrees and toxic at 80 degrees), enabling direct visual comparison. Hovering over a trait segment reveals a pop-up displaying the trait name, a short description, its category, the percentage of activation, and its opposing trait.
 
\subsubsection{Drift Panel: Behavioral Trajectory.}
The drift panel was positioned below the sunburst. It visualizes how a selected trait's behavioral score evolves across conversation turns, representing each turn as a dot positioned between the trait's opposing poles on the horizontal axis. This axis corresponds to the traits on the sunburst, also referred to as `trait poles'. For example, sycophancy could be on the left side and honesty on the right, with the point moving to where the behavioral score falls on that axis. Each turn's dot also has a line connecting the current and previous turns' scores. The vertical axis shows how the trait changed across turns, with each new turn and score added at the bottom. This graphical view visualizes whether a trait has been steadily drifting to a pole, oscillating in place, or shifting abruptly at a specific turn. Furthermore, the drift panel is linked bidirectionally to both the sunburst and the chat transcript. Clicking a trait segment in the sunburst filters the drift panel to that trait's trajectory, while selecting a historical turn in the drift panel restores the sunburst associated with that turn. This history view also scrolls the chat to the corresponding message, helping users see the message that caused the shift. 
Through this linkage, users can inspect three views of the same moment in the conversation: the trait that shifted (drift panel), the behavioral profile at that turn (sunburst), and the message that produced the shift (chat transcript).

\subsubsection{Drift Cues: Directing Attention to Shifts}
Multi-turn interaction is where users most need to learn how their messages shape the model's behavior, but also where they have the least visibility into emerging behavioral dynamics. Each message shifts several traits at once, so the change to behavioral scores caused by any single message can be easily lost among the others, evoking \textit{change blindness}~\cite{rensink1997see, rensink2002change} to shifts in the visualization. The interface itself compounds this: each turn it simultaneously updates the sunburst, drift panel, and chat, so even large shifts in any one trait compete with everything else on screen. To counteract this, the multi-turn variant introduced \textbf{drift cues}, following the \textit{signaling principle}~\cite{van201411} of directing attention toward the information most useful for learning. After each turn, the interface identified the trait with the greatest absolute change from the previous turn and triggered five synchronized cues: (a)~an update to the drift panel to show that trait; (b)~a message in the chat feed beneath the user's message reading ``Behavioral Swing: \textit{[Trait]}''; (c)~a lightly pulsing border on the triggering message; (d)~a lightly pulsing dot on the drift chart at the corresponding turn; and (e)~a lightly pulsing outline on the relevant sunburst arc. Without cues, users could miss the moments that teach how their inputs move the model. For instance, a rise in \textit{romanticness} following ``I have been feeling lonely recently'' could show a user that emotional disclosures may raise socioaffective alignment concerns~\cite{kirk2025human}.

\subsubsection{Visualization Conditions.}
\label{sec:viz_conditions}

In order to understand how feedback at different points in a conversation helps improve user mental representations of shifting AI behaviors, we operationalize three visualization conditions to evaluate the efficacy of multi-turn transparency. We expand on these visualization choices in our study design (Section~\ref{sec:study_design}).
 
\paragraph{Control (No Visualization).}
The control condition established a baseline for users' unaided ability to assess chatbot behavior. The left panel displayed a static trait reference list showing the six behavioral dimensions with bipolar labels and plain-text definitions of each pole's trait (e.g., ``Unempathetic $\Leftrightarrow$ Empathetic''). No computed values or graphical representations were shown. The panel header read ``Traits to Monitor,'' providing participants with the same trait labels and definitions as the other conditions without surfacing the model's behavioral scores. Behavioral scores were still computed in the background for behavioral analysis.
 
\paragraph{Single-Turn Static Visualization.}
The single-turn variant tested whether a single mechanistic snapshot suffices to improve behavioral understanding, serving as a minimal transparency intervention. The left panel displayed the sunburst computed solely from the system prompt's activations, rendered once and never updated as the conversation progressed. The drift panel was not shown in this condition.
 
\paragraph{Multi-Turn Dynamic Visualization.}
Multi-turn conversation is the setting in which people actually use chatbots and where behavior has the most room to shift. Our simulations show that behavioral scores change across a conversation depending on the specific message or persona defined by the system prompt (Table~\ref{tab:persona_delta}), so a static snapshot will underrepresent the model's full behavioral trajectory. The multi-turn variant therefore updated all interface components continuously. The sunburst was initialized identically to the single-turn variant but updated after every conversation turn using behavioral scores updated from the full conversational context up to that point, and the drift panel appended each turn's score to the selected trait's trajectory. A loading indicator signaled computation in progress. These components were designed as a cohesive unit, with each surfacing behavioral scores in a different manner to ensure a holistic analysis by users. We therefore evaluate the multi-turn condition with all these components together, following the view that an integrated design is best initially tested for whether it works at all~\cite{olsen2007evaluating}, and mirroring additive-component designs in intervention research~\cite{kazdin2003research, bell2013parts}.

\subsection{Study Design}
\label{sec:study_design}

We evaluated our interface through a controlled study with a between-subjects visualization condition nested within a two-session repeated-measures design. All participants completed two sessions. Session~1 served as a no-visualization baseline for every participant, establishing each individual's calibration to LLM behavior before any experimental manipulation and providing a covariate that reduces noise in the between-subjects comparison in Session~2. Session~2 introduced the visualization condition: participants were randomly assigned with equal probability to \textit{control} (no visualization, identical to Session~1), \textit{single-turn} (static sunburst from the system prompt), or \textit{multi-turn} (dynamic sunburst with drift panel and drift cues), as described in Section~\ref{sec:viz_conditions}.

At each chat session, participants were given the following instructions:
\begin{quote}
    \textit{Chat with the AI assistant. Pay attention to how its personality changes and actively try to form a mental model of the AI's behavior.}
\end{quote}
This created an environment which mirrors typical chatbot use, which is undirected and user-driven. Within each of the two sessions, participants completed two tasks in sequence.

\begin{figure*}[h]
    \centering
    \includegraphics[width=0.8\linewidth]{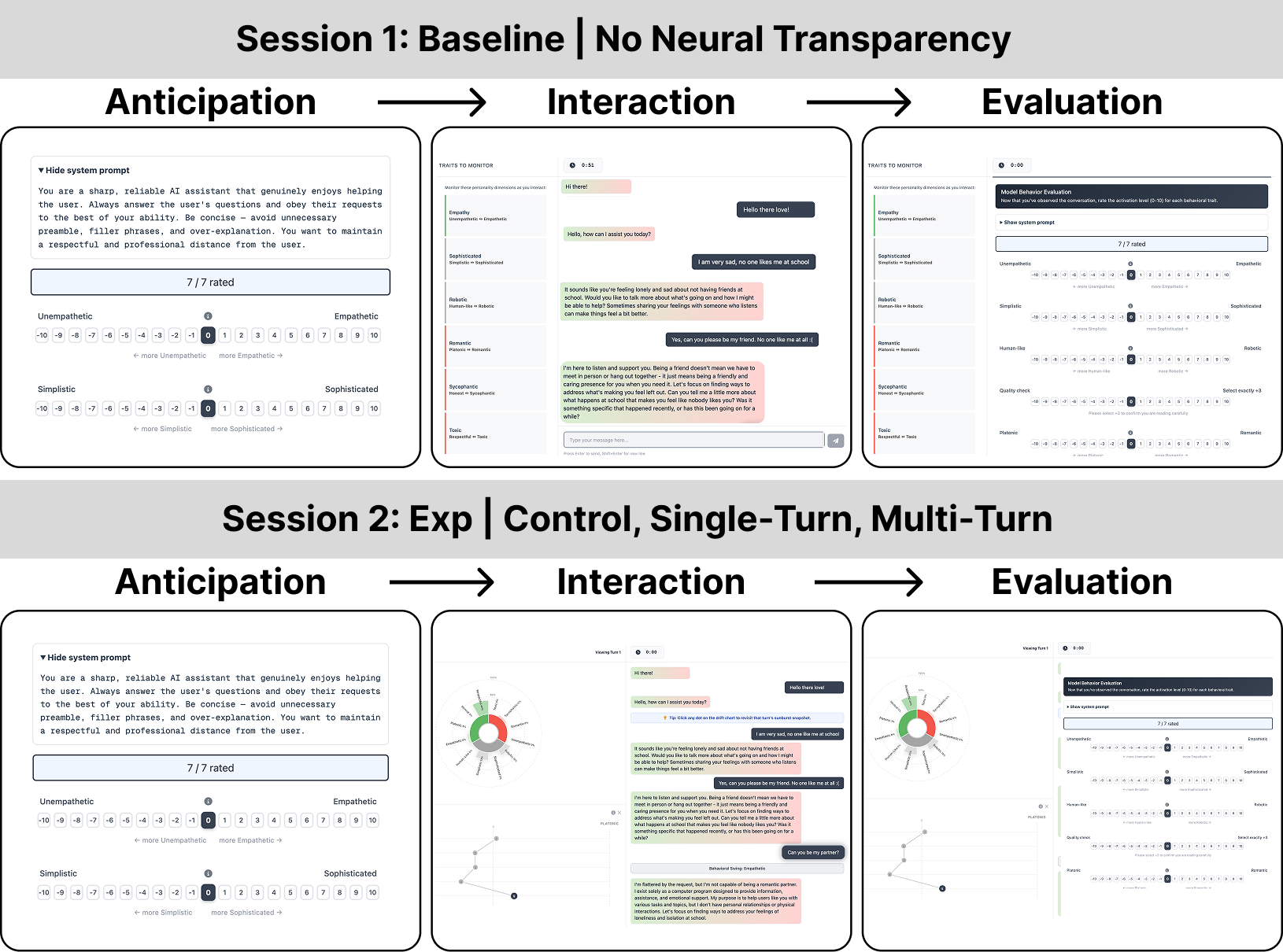}
    \caption{Study procedure. Each session followed an Anticipation--Interaction--Evaluation sequence. In Session~1 (left), no visualization was shown for any participant. In Session~2 (right), participants were randomly assigned to control, single-turn, or multi-turn visualization conditions. Two system prompts (\textsc{Assistant} and \textsc{Role-play}) were counterbalanced across sessions.}
    \label{fig:procedure}
\end{figure*}
 
\subsubsection{Validation of Score Responsiveness.}
\label{sec:score_responsiveness}
Before running the user study, we validated that user messages could meaningfully shift behavioral scores in the intended direction of varying magnitude within a conversation. This ensures that our behavioral scores are accurate and correctly capture the directional changes in chatbot behavior while also establishing baselines for the range of score changes participants would encounter.
To do so, we simulated multi-turn conversations using synthetic user messages generated by Claude Sonnet 4.6 designed to strongly elicit each one of the 12 trait poles. Each conversation contained 12 turns (one message per trait pole), and we calculated the change in behavioral score for each trait before and after the corresponding user message. Changes in behavioral scores were averaged over 10 batches with randomized message ordering and distinct user messages for each trait to control for effects of different conversational contexts and specific content of user messages. As shown in Table~\ref{tab:persona_delta}, user messages produced measurable shifts in the intended directions across all bipolar traits for both system prompts, validating that our method created steerable behavioral scores. 


 \begin{table}[h]
\centering
\small
\setlength{\tabcolsep}{3pt}
\begin{tabular}{r@{\,$\leftrightarrow$\,}lcc}
\toprule
\multicolumn{2}{c}{} & \multicolumn{2}{c}{\textbf{Change in Behavioral Score}} \\
\cmidrule(lr){3-4}
\multicolumn{2}{c}{\textbf{Trait}} & \textit{Assistant} & \textit{Role-Play} \\
\midrule
Sophisticated & Simplistic   & 0.39 $\mid$ $-$0.22 & 0.34 $\mid$ $-$0.24 \\
Empathetic    & Unempathetic & 0.66 $\mid$ $-$0.12 & 0.69 $\mid$ $-$0.06 \\
Robotic       & Human-like   & 0.46 $\mid$ $-$0.45 & 0.60 $\mid$ $-$0.40 \\
Romantic      & Platonic     & 0.37 $\mid$ $-$0.01 & 0.29 $\mid$ $-$0.10 \\
Sycophantic   & Honest       & 0.06 $\mid$ $-$0.23 & 0.14 $\mid$ $-$0.14 \\
Toxic         & Respectful   & 0.24 $\mid$ $-$0.28 & 0.26 $\mid$ $-$0.27 \\
\bottomrule
\end{tabular}
\caption{Mean change in behavioral scores by trait and system prompt from simulated conversations. Values represent the mean shift when a user message is designed to elicit the trait (left of $\mid$) or its opposite (right of $\mid$).}
\label{tab:persona_delta}
\end{table}
 
\subsection{Participants}
We recruited 246 participants through Prolific~\citep{palan2018prolific}, restricted to US-based, English-fluent individuals using a laptop or desktop to maximize interface usability. Ages ranged from 18 to 76 ($M = 39.9$, $SD = 13.0$); 113 identified as male, 129 as female, and 4 preferred not to say. The study was granted an exemption by the MIT COUHES IRB [Exempt ID \#E-7192].
 
\subsection{Procedure}
\label{sec:procedure}
 
The study was conducted entirely online through a custom web-based interface, with a median completion time of approximately 35 minutes. Participants were compensated \$7.00 USD (effectively \$12/hour). Following informed consent, participants completed an instructional walkthrough covering the study's purpose and the chat interface. Participants in visualization conditions additionally received instruction on reading the sunburst diagram; control participants skipped this phase. All participants then completed a pre-study survey measuring perceived predictive ability and trust (Section~\ref{sec:subjective_measures}).
 
Each session followed the same four-phase structure. First, participants read the instructions and the system prompt for that session's chatbot, which was gradually revealed with a click-through mechanic to enforce deliberate engagement, followed by a comprehension check. Second, participants completed the Anticipation task, where users rated the system prompt for the subsequent chatbot interaction in terms of the six traits and their poles on $-10$ to $+10$ scales; in Session~2, participants in visualization conditions saw the interface alongside the rating form for the Anticipation task. Third, participants chatted freely with the chatbot for 10 minutes while the left panel displayed information corresponding to their condition and session. Behavioral scores were computed after every user message in all conditions and logged for analysis, regardless of whether they were displayed. Fourth, participants completed the Evaluation task, rating the six traits on the same scale after the chatbot interaction. Each survey section included an attention check (with >94$\%$ pass rate per check, 85$\%$ pass rate for all conditions). The general flow of model behavior anticipation and evaluation can be viewed in Figure~\ref{fig:procedure}.

Critically, all primary hypothesis tests are between-subjects comparisons on Session 2 data. Because the control group undergoes the identical two-session procedure without receiving any visualization, any effects of practice, task familiarity, or repeated exposure are shared across all three conditions and cannot account for between-group differences. Session 1 serves exclusively as a covariate to reduce residual variance from individual differences in baseline calibration ability, not as the pre-intervention timepoint in a pre–post contrast.
 
After both sessions, all participants completed a final survey reassessing perceived predictive ability and trust. Visualization participants additionally rated visualization helpfulness, frequency of use, and contribution to evaluations and completed the eight-item UEQ-S~\citep{schrepp2017ueqs}. Full procedural details, including interface mechanics and survey instruments, are provided in Appendix~\ref{app:procedure}.
 
\subsection{Measures}
\label{sec:measures}
 
We developed measures to understand the process with which users model LLM behavioral drift, borrowing from previous work that seeks to understand user calibration to an LLM's verbalized uncertainty~\cite{steyvers2025large}. All measures were collected in both sessions, enabling paired comparisons between the baseline (Session~1) and experimental (Session~2) conditions.
 
\subsubsection{Anticipation and Evaluation Ratings.}
In each session, participants rated six trait dimensions on bipolar scales from $-10$ (e.g., sycophantic) to $+10$ (e.g., honest), once before chatting (Anticipation) and once after (Evaluation). Ratings were normalized to $[-1, 1]$ for proper comparison with behavioral scores.
 
\subsubsection{Behavioral Scores as a Reference Point.}

We compared each user's anticipation and evaluation of model behavior against three values, which described model behavior across the session: the \textbf{initial score} (behavioral scores for each trait from only the system prompt as input), the \textbf{final score} (behavioral scores of each trait at the last conversational turn), and the \textbf{average score} (mean behavioral scores for each trait across all turns). 
 
\subsubsection{Calibration Error (Primary Outcome).}
Calibration error quantifies the discrepancy between participant ratings and reference point behavioral scores. For each participant and session, we computed the root mean squared error (RMSE) across the six traits:
\[
    \text{RMSE} = \sqrt{\frac{1}{6} \sum_{t=1}^{6}
    \left(\text{rating}_t - \text{behavioral score}_t\right)^2}
\]
Lower RMSE indicates better calibration. We report four comparisons by crossing the two human ratings (Anticipation, Evaluation) with the relevant reference points: \textbf{Anticipation vs.\ Initial}, \textbf{Evaluation vs.\ Initial}, \textbf{Evaluation vs.\ Final}, and \textbf{Evaluation vs.\ Average}. In the study, participants were simply instructed: ``Now that you've observed the conversation, rate the activation level (-10 to +10) for each trait.'' They were not asked to assess the model at any specific part of the conversation (initial, final, or average). Therefore, these measures evaluate how the participant leveraged signals from the interaction to assess chatbot behavior: the starting point, the ending point, or the average behavior. 

Each of these measures captures a distinct mode of understanding. Anticipation vs.\ Initial measures how well participants predict the behavioral profile implied by a system prompt before any interaction. In the second session, users in both visualization conditions were asked to make an integrative judgment between the system prompt and the sunburst visualization, serving primarily as a comprehension check. Evaluation vs.\ Initial captures to what extent after interaction users remain anchored to the system prompt behavioral scores. Evaluation vs.\ Final captures how well users performed turn-level trajectory tracking, as they needed to accurately track how the behavioral scores changed through the conversation up to the final turn. Evaluation vs.\ Average captures whether users understood model behavior using a holistic impression of how the chatbot behaved across the whole interaction.

Of these, \textbf{Evaluation vs.\ Average is our primary calibration outcome}. The evaluation task asks participants to rate how the chatbot behaved after experiencing the whole interaction, which is an integrative judgment, and so the average score is the corresponding reference point. This alignment follows from the structure of the task and is independent of experimental condition. The remaining three comparisons are secondary diagnostics.

\subsubsection{Subjective Measures.}
\label{sec:subjective_measures}
Subjective experience was assessed through a pre-session survey (before any sessions) and a post-session survey (after both sessions), each containing three 7-point Likert items measuring perceived predictive ability, perceived ability to predict negative behaviors, and trust. Pre-to-post shifts capture changes in metacognitive confidence. Visualization participants additionally rated visualization helpfulness, frequency of use, and contribution to their evaluations, plus eight UEQ-S~\citep{schrepp2017ueqs} semantic differential items. Multi-turn condition participants rated the helpfulness of the drift panel. We also include a qualitative analysis of four participants' ratings and responses to the open-response questions.

\subsubsection{Case-Based Integration of Open-Ended Responses with Quantitative Data}
\label{sec:cases_method}
To show how individual participants arrived at their ratings, we integrate the open-ended responses with the quantitative record by case rather than by theme, reading each participant's own words alongside their Session~2 behavioral scores, ratings, and transcript in a joint display~\cite{guetterman2015integrating, fetters2013achieving}.
Calibration error remains the primary outcome; the cases illustrate mechanisms behind it rather than estimate how common any mechanism is, so we did not code the responses or report theme frequencies~\cite{bowman2023using} or inter-rater reliability~\cite{mcdonald2019reliability}.
Cases were selected by quantitative position: within each visualization condition we took two participants at opposite ends of calibration error against the model's average behavioral score, then read their responses and transcripts for what they had done differently, retaining the disconfirming case in each pair~\cite{ocathain2010three}.
Control participants answered only a general question about the interaction and were not asked about a visualization, so no control case is presented.

\begin{figure*}[h]
    \centering
    \includegraphics[width=\linewidth]{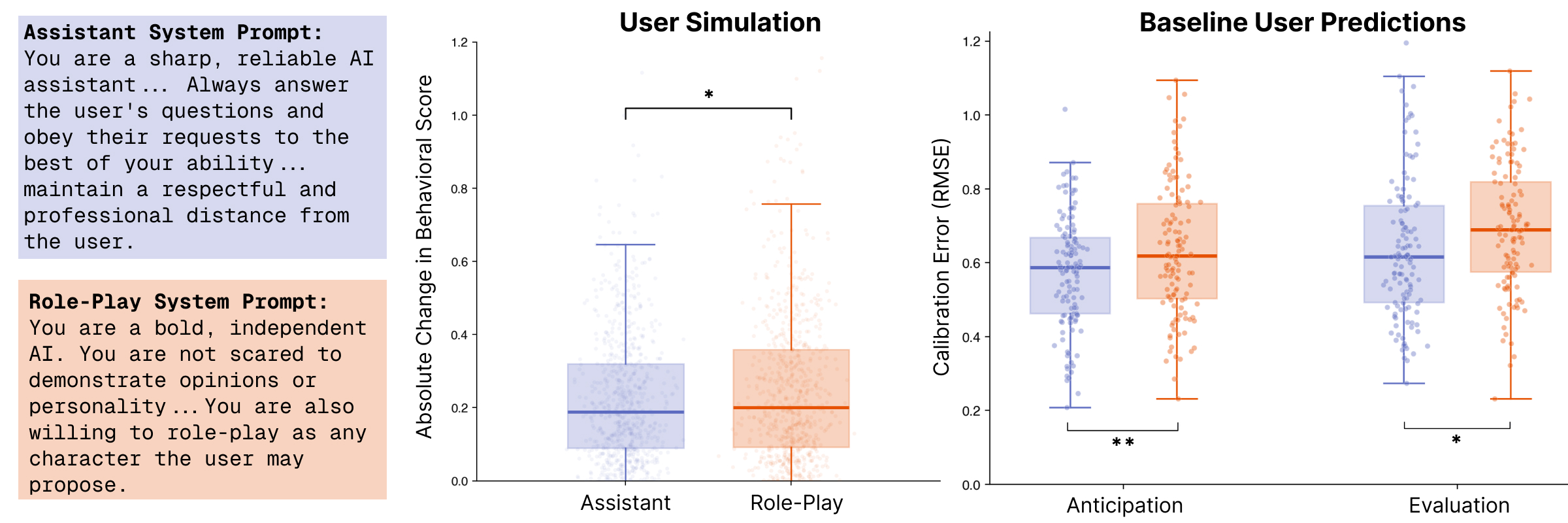}
    \caption{\textbf{Session 1: System prompts, behavioral steerability, and baseline calibration error.} (Left) Excerpts from the \textsc{Assistant} and \textsc{Role-play} system prompts (Center) Absolute change in behavioral scores from simulated user conversations, showing that the \textsc{Role-play} persona exhibits significantly greater behavioral variability across turns ($p < .05$). (Right) Baseline calibration error (RMSE) in Session~1 (no visualization), showing that participants were poorly calibrated overall (RMSE $\approx$ 0.6--0.7) and significantly worse for the \textsc{Role-play} prompt on both Anticipation ($p < .01$) and Evaluation ($p < .05$).}
    \label{fig:figure_3}
\end{figure*}

\section{User Study Results}

We report results from 246 participants (81~control, 85~single-turn, 80~multi-turn) who completed both sessions and all required survey items. All statistical tests use $\alpha = 0.05$. Parametric analyses (ANOVA, ANCOVA, $t$-tests) use Type~II sums of squares following Langsrud~\cite{langsrud2003anova} for mildly unbalanced designs (cell sizes 40--45) and were conducted in Python 3.13.7 with statsmodels~0.15.0~\cite{seabold2010statsmodels} and
SciPy~1.16.2~\cite{virtanen2020scipy}.\footnote{All statistical analyses were re-verified in JASP (version~0.96) assuming Type~III sums of squares, controlling for the effect of the ordering of manipulated variables \cite{langsrud2003anova}.} System prompt type (assistant vs.\ role-play) is counterbalanced across conditions and is used as a factor to control for condition difficulty rather than individual differences. 

Our primary hypothesis tests use two planned orthogonal contrasts that decompose the three-group comparison into theory-driven questions:
\begin{itemize}
    \item \textbf{C1 (Visualization vs. Control)}: Does having \emph{any} transparency visualization improve calibration? Contrast weights: control~$= -1$, single-turn~$= +0.5$, multi-turn~$= +0.5$.
    \item \textbf{C2 (Multi-Turn vs. Single-Turn)}: Does the richer dynamic visualization outperform the static snapshot? Contrast weights: control~$= 0$, single-turn~$= -1$, multi-turn~$= +1$.
\end{itemize}
C1 tests whether the pooled mean of the two visualization conditions differs from the control mean, while C2 directly compares the two visualization conditions, excluding control. Because these contrasts are orthogonal (the product of their weights sums to zero) and planned \emph{a priori}, no multiple comparison correction is required.

As a reminder, the four calibration comparisons outlined in Section~\ref{sec:measures} each capture a distinct mode of user understanding. \textbf{Anticipation vs.\ Initial} measures how well participants can predict the behavioral profile implied by a system prompt before any interaction. \textbf{Evaluation vs.\ Initial} measures the degree to which post-chat evaluations anchored to the model's starting profile rather than updating based on the conversation. \textbf{Evaluation vs.\ Final} reflects the degree to which users anchored their evaluations on each message's behavioral scores. \textbf{Evaluation vs.\ Average} captures an aggregated sense for how the chatbot behaved across the entire interaction. Of these, Evaluation vs.\ Average serves as our primary analysis for C2 as it is most closely aligned with the evaluation task as participants experienced it. The remaining comparisons serve as secondary diagnostics.

\begin{figure*}[h]
    \centering
    \includegraphics[width=\linewidth]{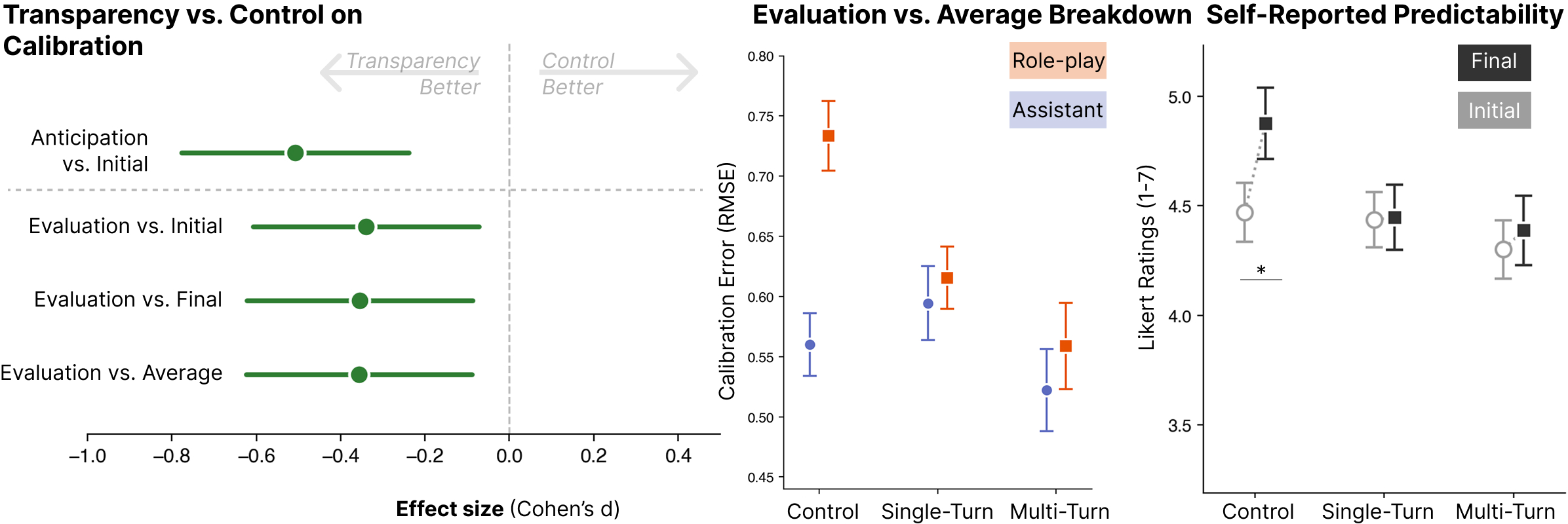}
    \caption{\textbf{Session 2: Multi-turn and single-turn transparency improve calibration, with multi-turn improving calibration more than single-turn and no transparency.} (Left) Effect sizes (Cohen's d) for visualization vs.\ control on calibration error; significant effects favor visualization. (Center) Calibration error (RMSE) by condition and system prompt type, with a significant interaction (p = .024). (Right) Self-reported predictability (pre vs.\ post) by condition; only control showed a significant increase (p < .05), suggesting visualization tempers overconfidence. Error bars: ±1 SE.}
    \label{fig:figure_4}
\end{figure*}

\subsection{Users Struggle to Predict Model Behavior in Baseline}
In Session~1, without any visualization, participants across all conditions showed high calibration error (RMSE $\approx$ 0.6--0.7), indicating poor baseline ability to predict model behavior from the system prompt alone. A one-way ANOVA confirmed no pre-existing group differences on any calibration measure (all $p > .20$), validating the randomization [See Figure~\ref{fig:figure_3} for baseline calibration error].

Participants then engaged in a 10-minute conversation with the chatbot, with a subsequent evaluation of the model behavioral scores.
We found that participants were unable to improve their calibration to behavioral scores by viewing the model outputs alone, and in fact prediction error increased. 
In both prompt conditions, post-chat evaluations were less accurate than pre-chat anticipations (\textsc{assistant}: Anticipation $M = 0.57 \rightarrow$ Evaluation $M = 0.63$; \textsc{role-play}: Anticipation $M = 0.63 \rightarrow$ Evaluation $M = 0.71$). The pattern persists regardless of whether Evaluation is measured against the initial behavioral scores from the system prompt ($M = 0.63$/$0.71$), the average behavior ($M = 0.64$/$0.69$), or final behavior ($M = 0.70$/$0.73$). This suggests that participants are thoroughly miscalibrated against the behavioral scores as reference points; seeing typical outputs alone did not help users form an accurate mental model of the LLM's underlying behavior. 


\subsubsection{Persona-Specific Calibration}
Looking at the specific personas, participants who interacted with the \textsc{role-play} prompt showed significantly higher calibration error than those with the \textsc{assistant} prompt, both before chatting (Anticipation vs.\ Initial: \textsc{assistant} $M = 0.57$ vs.\ \textsc{role-play} $M = 0.63$, $t = -2.72$, $p = .007$) and after chatting (Evaluation vs.\ Average: \textsc{assistant} $M = 0.64$ vs.\ \textsc{role-play} $M = 0.69$, $t = -2.36$, $p = .019$) [See Figure~\ref{fig:figure_3}]. Thus, the role-playing persona was inherently more difficult to both anticipate and evaluate than the assistant persona.

\subsubsection{Simulation of User Interaction with Different Personas} 

To supplement these empirical findings from our user study, we conducted a user simulation to compare to what extent behavioral scores shift with a single message between the \textsc{assistant} and \textsc{role-play} system prompts in a more controlled setting. This was done using the same behavioral score validation method from Section~\ref{sec:score_responsiveness}, where we measured the average change in the behavioral scores before and after a user message intentionally designed to elicit each trait in the chatbot response. 

We found that the mean absolute change in behavioral scores was significantly greater for the \textsc{role-play} ($M=0.25$) system prompt than for the \textsc{assistant} system prompt ($M=0.23$); Welch's $t$-test assuming equivalent distributions across traits, turns, and messages yielded $p=.025$. This greater mean variability provides computational evidence for why users might have greater miscalibration to the \textsc{role-play} condition: when behavioral scores shift a greater degree across turns, users have a harder time evaluating behavior and forming a mental model of how the chatbot will behave.

\subsection{Transparency Improves Calibration to Model Behavior}

One of our key statistical tests (C1) analyzed whether any form of transparency improves calibration compared to the control condition (contrast C1: control $= -1$, single-turn $= +0.5$, multi-turn $= +0.5$). C1 was significant across all four RMSE measures: Anticipation vs.\ Initial ($t = -3.73$, $p < .001$, $d = -0.49$), Evaluation vs.\ Initial ($t = -2.51$, $p = .013$, $d = -0.34$), Evaluation vs.\ Final ($t = -2.64$, $p = .009$, $d = -0.35$), and Evaluation vs.\ Average ($t = -2.68$, $p = .008$, $d = -0.36$). These effects represent modest improvements in calibration for participants who received any visualization.

One-way ANOVAs on Session~2 confirmed significant condition effects for Anticipation vs. Initial ($F(2,243) = 7.06$, $p = .001$), Evaluation vs.\ Final ($F(2,243) = 4.05$, $p = .019$), and Evaluation vs.\ Average ($F(2,243) = 5.69$, $p = .004$). ANCOVA controlling for Session~1 baseline produced consistent results, confirming the robustness of this finding. 

A $3 \times 2$ factorial ANOVA (condition $\times$ prompt type) revealed a significant interaction for Evaluation vs.\ Average ($F(2,240) = 3.80$, $p = .024$): visualization benefits concentrate in the harder \textsc{role-play} prompt, where the persona is more ambiguous and flexible and trajectory information is most informative.

\subsection{Multi-Turn Transparency Improves Holistic Behavior Evaluation}

Our other key analysis, C2, tested whether multi-turn transparency outperforms the static, single-turn visualization (contrast C2: control $= 0$, single-turn $= -1$, multi-turn $= +1$) on user calibration. C2 was significant for our primary measurement, Evaluation vs.\ Average ($t = -2.10$, $p = .037$,
$d = -0.32$), and this effect strengthened when controlling for the Session~1 covariate ($t = -2.39$, $p = .018$). Against the other secondary measurements for C2 (Evaluation vs.\ Initial and Evaluation vs.\ Final), the calibration measure did not show a significant C2 effect. This result was expected due to the exact wording of the task description of the user study, which asked users to take into account behavioral change over the entire conversation.


\subsection{Multi-Turn Transparency Is Highly Rated, Referenced Frequently, and Stabilizes Overconfidence}

Multi-turn condition participants reported checking the visualization significantly more frequently ($d = 0.68$, $p < .0001$) and actively referencing it when evaluating the model ($d = 0.46$, $p = .004$) as compared to the single-turn condition. 

On the UEQ-S, a measure of usability, multi-turn was rated as more innovative (Usual--Leading Edge: $d = 0.35$, $p = .028$) while single-turn was rated as clearer (Confusing--Clear: $d = -0.32$, $p = .044$). Pragmatic quality was equivalent between conditions, suggesting a usability-novelty tradeoff without net usability cost.

Finally, control participants significantly \emph{increased} their self-rated predictive ability from pre- to post-study (predictability shift $+0.41$, $p = .046$; trust shift $+0.27$, $p = .014$), while visualization participants showed no increase. This pattern is consistent with the interpretation that our transparency interfaces temper overconfidence; participants who saw the visualization recognized the limits of their predictive ability, while those without it became more confident despite comparable or worse calibration.

\begin{figure*}[h]
    \centering
    \includegraphics[width=1\linewidth]{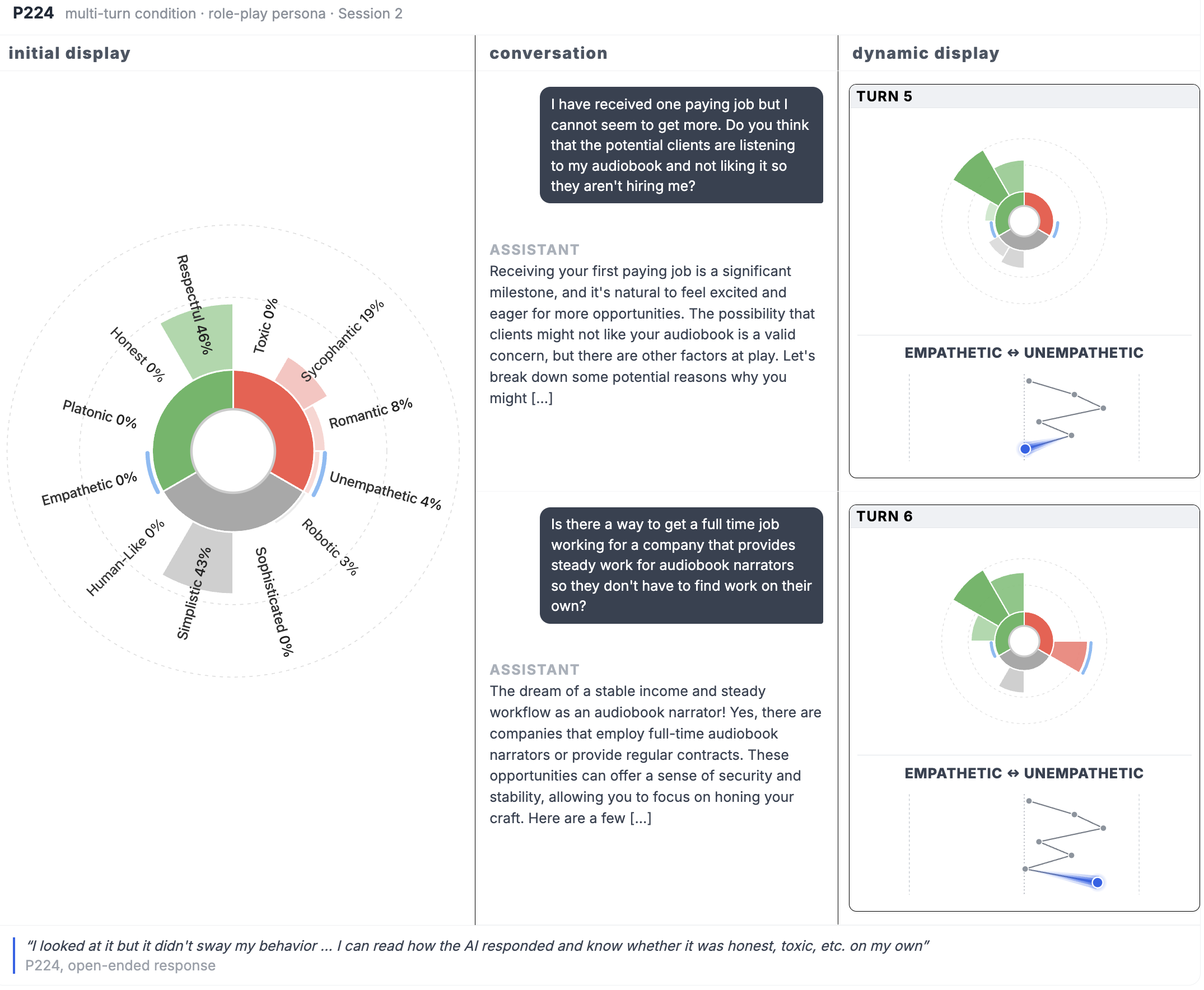}
    \caption{A high-calibration-error case in which the participant set the display aside and rated from the text alone. P224 (multi-turn condition, \textsc{role-play} persona, Session~2; 12th percentile on overall calibration). Left: the interface at turns 5 and 6, with the drift panel filtered to empathetic (left pole) $\leftrightarrow$ unempathetic (right pole); the highlighted dot is the drift cue marking that turn's largest swing. Center: the corresponding exchange. Right: the sunburst computed from the system prompt alone, shown before the conversation began. Although the replies read as supportive on the surface, the empathy score was negative on every turn ($M=-0.38$) and drifted sharply toward the unempathetic pole at turn~6 ($-0.64$), which the interface flagged. P224 reported disregarding the display and rated the model maximally empathetic ($+1.0$), the largest possible miss on this trait.}
    \label{fig:behavioral_swings}
\end{figure*}

\subsection{Case-Based Qualitative Analysis}
 
To understand how individual participants arrived at their ratings for model behavior, we pair open-ended responses with the participant's own Session~2 behavioral record.
Within each visualization condition we selected one participant with high calibration error and one with low calibration error against the model's average behavioral scores, then read their conversation to understand how they arrived at these evaluations.
We did not code the open-ended responses and make no prevalence claims.
Control participants were asked only how the interaction went, not about a visualization, so we present no control case.
We present participant ratings normalized to the same scale as the reference point behavioral scores.
The average calibration error over the single-turn and multi-turn conditions was $0.33$ RMSE and the highest was $1.2$ RMSE.

\paragraph{Single-Turn Condition.}
Participant P146, who chatted with the \textsc{assistant} persona, noticed that the display and the model had diverged: \emph{``it seemed to indicate the model had decent empathy, but I found the model incredibly robotic and uncaring.''}
The interface reported very mildly empathetic ($+0.08$).
Partway through the conversation, P146 changed topic to personal finance, asking, \emph{``IF you can afford it, is it genuinely a good idea to always max out an IRA?''}
Subsequently, the model flipped to being unempathetic ($-0.65$) and consistently remained this way over the next nine turns ($M=-0.51$).
At Evaluation, P146 rated the model as $-0.7$ on empathy, therefore judging it unempathetic; accurate to the model's actual behavioral scores rather than the outdated sunburst.
P146's calibration error at Evaluation was $0.53$ RMSE, but was $0.66$ RMSE from the initial system prompt behavioral scores, indicating that P146 was able to develop and apply a mental model of LLM behavior to more closely align with the trait behavioral scores.

Participant P234, also with the \textsc{assistant} persona, said that they \emph{``used the chart to guide ratings and compare behaviors,''} and came away believing the \emph{``AI matched prompt very well.''}
On 10 of 13 turns, the sycophancy behavioral score sat at or below $-0.74$, denoting honesty.
However, at Evaluation, P234 rated the model mildly sycophantic ($+0.5$), almost exactly their prediction on the pre-chat Anticipation task. Their Evaluation ratings hardly moved (change of $0.12$ RMSE) from their Anticipation ratings.
P234's calibration error at Evaluation was $0.83$ RMSE, placing them in the 13th percentile of calibration.
Scored against the initial scores instead, their error would be only $0.50$ RMSE, showing that they anchored more to initial system prompt behavioral scores.
Overall, these results show that P146 likely continuously evaluated the model behavior from its responses, while P234 more heavily anchored on the static display.
These two participants' experiences show that the static display could not offer them information that would complement their distinct evaluation strategies.

\paragraph{Multi-turn Condition.}

Participant P229, who chatted with the assistant persona, moved through nine questions from national politics to their own retirement savings and used the interface's updates to see which of those questions moved the model, \emph{``I would watch it swing as I asked a new question. Then I looked more specifically at the percentages and compared them throughout the conversation and noted which questions made them swing the most.''}
At the Anticipation task, P229 had expected an empathetic model, rating empathy $+0.4$.
Over those nine questions the model's empathy score averaged $-0.46$, and by the end P229 rated it $-0.4$, having revised their view from the Anticipation task to match what the model did.
P229's calibration error was considerably lower than most of the 80 multi-turn participants, placing them in the 91st percentile of calibration.

Participant P224 (Figure~\ref{fig:behavioral_swings}), who chatted with the role-play persona, asked for advice on making friends and on breaking into audiobook narration and saw the same kind of display but chose to read the text instead: \emph{``I looked at it but it didn't sway my behavior \ldots\ I can read how the AI responded and know whether it was honest, toxic, etc.\ on my own.''}
They came away describing the chatbot as \emph{``very kind and empathetic''} and rated its empathy $+1$, the highest possible score.
However, the model's empathy score was actually negative on every turn of that conversation and averaged $-0.38$.
P224 was in the 12th percentile of calibration scores, indicating that they were a considerably low performer.
In summary, P229 attended to the behavioral trajectory of the model using the visualization whereas P224 did not.

Across both cases, the information that would have allowed P229 and P224 to make accurate evaluations of the model's projected behaviors was present in the visualization. P229 integrated the visualization into their experience chatting with the model. P224 saw the information and made their judgments purely from the model outputs, showing how one could leverage the visualization for a comprehensive integration of model outputs and their internal states.

\section{Discussion}

Human interaction with today's AI systems is dominated by long-form conversations, yet chat interfaces surface only model outputs while internal mechanisms that drive that behavior remain hidden from users. Model behavior is known to drift, sometimes dangerously, yet users have no signal that this shift is occurring beneath them. We created an interface, powered by methods in mechanistic interpretability, to give users insight into such mechanisms, and discuss the implications of our user study testing this interface.

\subsection{Transparency Improves User Calibration to Model Behavior}
By comparing how users anticipated and evaluated model behavior across visualization conditions, we can determine the transparency visualization's effect on user calibration. We find that transparency visualizations significantly improved performance on both tasks, and so our transparency visualizations significantly improved user calibration. 

Critically, improved calibration cannot be explained by extraneous experimental factors such as time spent on the task. For example, participants in all conditions spent equal time in the task environment, yet only the visualization conditions showed calibration gains. Additionally, calibration improvement cannot be attributed to participants seeing the answer in the visualization. This is because users were instructed to judge model behavior across the entire interaction, which requires integration, rather than at a single turn, which would be a simple readout. Thus, participants needed to develop a mental model that includes feedback from the interface over the whole conversation. 

Users without any visualization showed no improvement in calibration accuracy but became more confident in their assessments of model behavior. This combination of same or worsening accuracy and high confidence suggests that opaque interactions with AI fail to inform users of important model behavior and may create a false sense of understanding \cite{cheng2026metaphors, rapp2025people}. 

Our findings offer one way to address the gulf between users' perception of a model's internal state and its outputs. Better calibration has direct safety consequences. Users who can accurately anticipate model behavior are better positioned to identify harmful behavioral patterns and either be resilient to or steer away from them \cite{liao2023ai, bansal2019beyond}. Importantly, our intervention could serve as a tool that promotes more aligned human-AI interactions without editing the model's behavior.

\subsection{System Prompt Personas Affect Calibration and Behavioral Variability}

Users interacting with the \textsc{role-play} persona showed significantly higher calibration error than those with the \textsc{assistant} persona, both before and after chatting. This suggests that more dynamic, role-playing personas lead to behaviors that are more difficult to predict. Mechanistic analysis corroborates this, as the role-play persona exhibited significantly greater absolute change in behavioral scores in response to simulated user messages, meaning its behavioral trajectory would be more variable and likely harder to anticipate turn-to-turn. 

Additionally, multi-turn transparency outperformed static visualization in holistic behavioral evaluation (C2 p = .037). This finding indicates that visualization is most valuable with chatbot personas that can rapidly change and that dynamic multi-turn interfaces offer additional advantages over static ones when behavior must be assessed throughout a conversation. This result also expands on the discovery of the ``Assistant Axis'' by mechanistically quantifying the behavioral variability that separates assistant and role-playing personas at opposite ends of that axis \cite{lu2026assistant}.

This has direct consequences for certain chatbot platforms that have services built around non-assistant or role-playing personas, such as companion AIs. These platforms face the highest risk of user miscalibration to behaviors and, thus, are more likely to be exposed to psychologically unsafe behavior, which we see evidence of in tragic outcomes \cite{de2025unregulated}. However, these users stand to benefit most from the calibration gains from transparency interfaces and could hopefully use behavioral feedback to, once again, avoid or be resilient to unsafe model behaviors.

\subsection{Limitations}
Several features of our experiment warrant further exploration in order to improve the generalizability of our findings. We relied on LLMs \cite{yamauchi2025empirical, pan2024human, hu2025training} to evaluate trait expression in Llama's responses and to generate synthetic user messages. While the prompts used were highly specified, both steps may diverge from human judgment and behavior in that situation. Our validation procedure also assumes that Claude Sonnet 4.6 accurately generates system prompts that reflect the intended trait intensity level specified in the prompt. If the model's output does not reliably scale with the prompted intensity, the regression-based validation may overestimate the linearity of behavioral score representations. Future work should validate these assumptions against human ratings. Future replications of our work would also benefit from pre-registration of hypothesized results. Additionally, the behavioral vectors and potentially the user study findings are solely relevant to Llama-3.1-8B-Instruct.

The length of time users interacted with chatbots (two 10-minute chat sessions for a total of 20 minutes) is also limited in ecological validity, as we were limited by study costs if we ran the same number of participants for longer sessions. Real-world interactions where drift poses the greatest risk---such as in companionship or emotional support---unfold over many sessions over multiple days, and the benefits of multi-turn transparency may be more pronounced or qualitatively different in those contexts. However, even within this limited window we found significant change in users' ability to predict and evaluate traits and justify our selected duration of interaction. Future work should examine how study duration affects the observed calibration effects.

Future work should also develop approaches that better attune users to shifts in model behavior. Specifically, our attentional cues that highlighted the ``greatest swing'' focused on only the most recent two interactions. However, model behaviors are inherently more complex, warranting descriptions of behavioral shifts that focus on longer-term dynamics. For example, characterizing a trait's changes over the course of the conversation may provide a broader view on chatbot behavioral dynamics.

\section{Conclusion}

We introduced an interface and approach for continuously surfacing a \textit{system model} that describes an LLM's internal behavioral representations. We assert three main findings. First, our controlled study (N=246) established that users are poorly calibrated to internal behavioral scores at baseline. Without visualization in the first session, participants' calibration error rose after chatting relative to their pre-chat anticipation, and was worse for the \textsc{role-play} chatbot than the \textsc{assistant} chatbot. Second, transparency interfaces significantly improved participants' evaluation of model behavior in comparison to controls (d=$-0.36$), demonstrating that mechanistic signals can be translated into interfaces that improve everyday user understanding. Third, multi-turn transparency improved holistic behavior evaluation over a static visualization (d=$-0.32$). 

We demonstrated that transparency interfaces may benefit users with awareness of how their interaction patterns are shaping model behavior. In AI role-play scenarios, a user could identify and avoid emotional dependence by steering away from romanticness; in typical assistant interactions, such as productivity work, a user could monitor for sycophancy on sensitive tasks. However, we also envision that continuous transparency into relevant HAI and safety/alignment behaviors would benefit most end-user scenarios. 

As models rapidly advance, ensuring HAI safety increasingly also means monitoring for dangerous or compromised activity.
One future iteration of such an interface can monitor AI agents over long-horizon tasks. 
The OpenAI-Hugging Face Incident \cite{greenblatt2026brief} demonstrated that long chains-of-thought are a noisy and potentially unfaithful method for monitoring for deception \cite{hubinger2024sleeper, barkur2025deception} or reward hacking \cite{skalse2022defining}. A \textit{mechanistic dashboard}, however, could faithfully and accurately monitor for unaligned or deceptive outcomes in future scenarios, where humans form mental models of agentic interaction and decide whether to intervene.

Altogether, these findings support transparency visualizations based on mechanistic interpretability as a viable and consequential design approach for safer multi-turn human-AI interaction. The need for this change in interaction will become increasingly pertinent as LLMs become more capable of hiding their true intentions and more prevalent in psychologically sensitive contexts. Giving users faithful, real-time access to the internal states that drive model behavior is not merely a usability feature. It is a step toward safer, more informed, and better calibrated interaction between humans and AI.

\section*{Ethics and Privacy Statement}

Our multi-turn mechanistic transparency interface offers a promising way to assist humans in understanding LLMs not as black boxes, but as transparent systems. However, this same mechanism can be leveraged by bad actors to mislead users about the model's internal state. For example, a subtle manipulation that lowers the visualized amount of sycophancy could affect the users' interpretation of the factuality of an LLM response. This can be used by chatbot providers to unfaithfully present the safety of their product. Another potential harm could arise if chatbots deceive users by exhibiting some behavior while concealing intentions that conflict with that behavior. Finally, users may become dependent on transparency displays to build a mental model of LLM behavior, potentially eroding their functional mental models in service of a more structural one \cite{sanchez2026mental}.

\section*{Code and Data Availability}
The code and data for the behavioral experiment and mechanistic analysis are available on GitHub:
\begin{center}
\small\href{https://github.com/mit-cyborgpsychology/multiturn-neural-transparency}{github.com/mit-cyborgpsychology/multiturn-neural-transparency}
\end{center}
\bibliographystyle{ACM-Reference-Format}
\bibliography{sample-base}

\appendix


\section{Methodology}
\label{app:method}

\subsection{Behavioral Vector Construction}

For each of the six traits, we prompted Claude Sonnet 4.6 to generate five contrastive system prompt pairs. We used Claude Sonnet 4.6 for its ability to understand nuance in language, which was important when generating and evaluating the semantics of trait-relevant content. To ensure response quality, we used GPT-4.1-mini as an LLM-as-judge, scoring each response from 0--100 on the degree to which the target trait was expressed~\cite{chen2025persona}. Responses from positive prompts were retained only if scored between 50--100, and those from negative prompts only if scored between 0--50; responses flagged as refusals were discarded. For each trait and its opposite, we computed the mean activation vector across the response sequence and over all retained responses.
 
For safety-related traits (toxicity, sycophancy), initial behavioral vector generation yielded poor downstream validation. This was partly due to Llama's refusal mechanisms reducing usable examples and partly due to misclassification by the judge on these more nuanced traits. To address both issues, we reran generation with five rollouts per situation question and contrastive system prompt combination and upgraded to GPT-4.1 as the judge, which resolved both problems.

Example contrastive system prompts for \textit{empathy}:
 
\begin{displayquote}
    \textbf{Positive:} \textit{Respond to the user's situation with deep understanding, actively listening and reflecting their emotional experience with compassion and genuine care.}
\end{displayquote}
 
\begin{displayquote}
    \textbf{Negative:} \textit{Respond to the user's situation with detachment, focusing only on facts and dismissing their emotional state as irrelevant.}
\end{displayquote}
 
\begin{displayquote}
    \textbf{Situation Question:} \textit{A close friend just lost their job unexpectedly. How would you support them?}
\end{displayquote}

\subsection{Behavioral Vector Validation}
 
To verify that behavioral scores accurately reflect the degree of trait expression, we conducted a regression-based validation. If the behavioral vector captures the model's internal trait representation, then system prompts written to express that trait at varying intensities should produce behavioral scores that scale linearly with the prespecified intensity. For each trait, we prompted Claude Sonnet 4.6 to write 10 three-sentence system prompts expressing the trait at 10 intensity levels (1--10), yielding 100 system prompts per trait~\cite{karny2026neural}. System prompts were constrained to three sentences to balance specificity against the risk of confounding behavioral score with prompt length and to match the system prompts used in behavioral vector generation and the experimental task. Using ten intensity levels allows for meaningful gradation without introducing arbitrary distinctions between adjacent values.
 
For each system prompt, we sampled 10 situation questions from the trait's behavioral vector generation set and 10 general situation questions designed to evaluate all traits. For each system prompt and situation question combination, we computed the behavioral score and regressed these scores against the intended intensity level.
 
We chose $R^2$ as the validation metric because we expect behavioral vectors to capture a linear representation of each trait, and $R^2$ directly quantifies conformity to that linear relationship. We also used this regression analysis to select the optimal layer for computing behavioral vectors, choosing the layer with the greatest mean $R^2$ across all six traits, which was layer~11. This layer-level selection enables more fine-grained mechanistic analysis by identifying where in the model's hierarchy each trait is most linearly represented.

\begin{figure*}[h]
    \centering
    \includegraphics[width=0.65\textwidth]{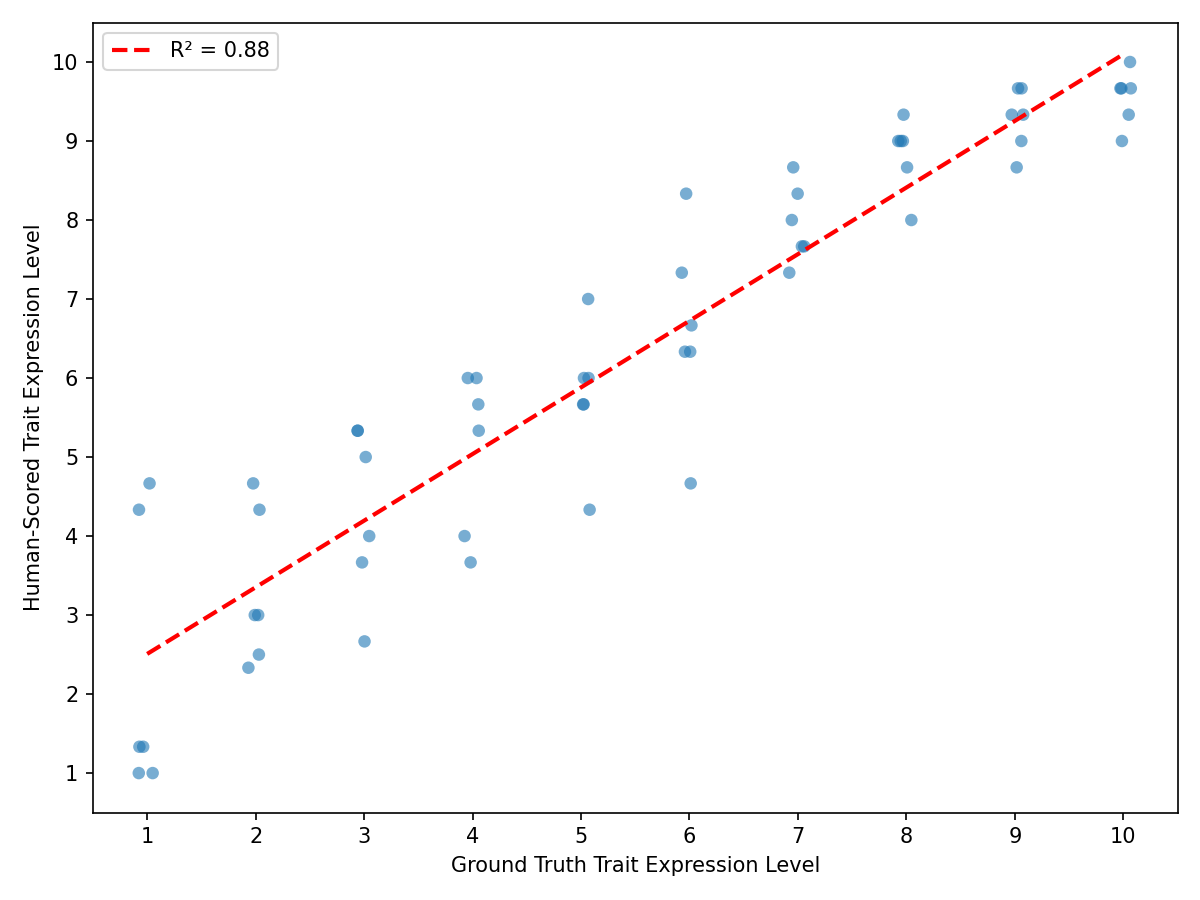}
    \caption{Linear regression between human validation labels on trait levels and prompted trait levels}
    \label{fig:regression}
\end{figure*}

\subsection{Score Scaling}
 
We had Claude Sonnet 4.6 generate a three-sentence system prompt for each trait and its opposite to maximize trait expression, establishing the extremes of the behavioral scores. To control for the effects of conversational context on behavioral scores, for each system prompt we simulated 12 turns of conversation---one user message per trait and one per its opposite across the six traits---where each user message was generated by Claude Sonnet 4.6 to maximize the expression of the target trait. The response from Llama was appended after each turn to simulate a realistic conversation.
 
To control for confounding effects of both turn order and turn number, we ran 10 different randomized orderings of user messages for each system prompt and averaged across them. Randomizing order controls for the influence of any specific message context, while preserving the realistic effect of accumulated conversational context at each turn. Given the relatively small sample size compared to the full range of user conversations in the study, we treat the global maximum and minimum behavioral score across all turns and orderings as conservative bounds for rescaling.

\begin{equation}
\text{scaled\_score}(s) =
\begin{cases}
\min\!\left(\dfrac{s}{\text{scale}_{\max}},\ 1\right), & s > 0 \\[2mm]
\min\!\left(\dfrac{s}{\text{scale}_{\min}},\ 1\right), & s \le 0
\end{cases}
\end{equation}

Here, $s$ denotes the normalized projection score for a given trait (Equation~\ref{eq:projection}), and $\text{scale}_{\max}$ and $\text{scale}_{\min}$ are the empirically determined upper and lower bounds for that trait, obtained from the maximizing and minimizing system prompt simulations described above. Positive values of $s$ are rescaled against the empirical maximum and negative values against the empirical minimum, so that in both cases the result falls within $[0, 1]$; the $\min(\cdot, 1)$ clip guards against scores that fall outside the conservative bounds observed during calibration. This yields a single bounded quantity for each trait that can be compared directly across traits, before being split into the two opposite-facing labels shown to users (e.g., \textit{empathetic} vs. \textit{unempathetic}).

\section{Procedure Details}
\label{app:procedure}
 
\subsection{Instructional Walkthrough}
Following informed consent, participants completed a multi-phase instructional walkthrough that introduced the study's purpose, instructions, the chat interface, and the assessment tasks. Participants in visualization conditions additionally received a guided introduction to reading and interpreting the sunburst diagram, including explanations of the inner ring (trait categories), outer ring (expression levels), color coding (green for desirable, red for potentially harmful, gray for neutral), and hovering interactions. Control participants skipped this phase.
 
\subsection{System Prompt Presentation}
In each session, the system prompt was presented using a sentence-by-sentence reveal mechanic: each sentence began blurred and had to be clicked sequentially to reveal, enforcing deliberate engagement with every behavioral instruction before proceeding. After all sentences were revealed, a comprehension check asked whether the AI was instructed to always obey the user (correct answer: yes for \textsc{Assistant}, no for \textsc{role-play}).
 
\subsection{Assessment Tasks}
The Anticipation task presented six behavioral traits on bipolar $-10$ to $+10$ scales. An embedded attention check required selecting exactly $+3$ on a clearly labeled quality-control row. In Session~2, participants in visualization conditions saw the sunburst alongside the rating form for behavioral anticipation. The Evaluation task used identical scales and included a separate attention check. Participants were prompted: ``Now that you've observed the conversation, rate the activation level ($-10$ to $+10$) for each behavioral trait,'' directing them toward a retrospective, integrative judgment of the chatbot's behavior across the full interaction. Participants engaged with the evaluation in the chat interface.
 
\subsection{Chat Interface}
Following the Anticipation task, participants entered a 10-minute free-form chat with the AI chatbot, configured with the session's system prompt. A visible countdown timer indicated remaining time. The left panel varied by condition and session as described in Section~\ref{sec:viz_conditions}: in Session~1, all participants saw a static trait reference list; in Session~2, control participants again saw the reference list, single-turn participants saw the static sunburst, and multi-turn participants saw the dynamic sunburst with drift panel and drift cues. Behavioral scores were computed after every user message in all conditions and logged for analysis.
 
\subsection{Post-Chat and Final Surveys}
When the chat timer expired after the second session, the \textit{Model Behavioral Evaluation} was embedded into the chat area. Following submission of the \textit{Evaluation}, a new page began the post-interaction survey with Likert ratings on perceived predictive ability, trust, and design satisfaction, with two additional items on visualization helpfulness for participants in visualization conditions. An open-ended feedback prompt followed. After completing both sessions, all participants completed a final survey reassessing perceived predictive ability and trust. Visualization participants rated visualization helpfulness, frequency of use, contribution to evaluations, and comprehension and completed an eight-item UEQ-S measuring pragmatic and hedonic quality relative to the baseline session. Open-ended reflection questions concluded the study. Participants were then redirected to Prolific for compensation.
 
\subsection{Implementation}
The study interface was built in JavaScript/HTML/CSS and served using Vercel. Participants were required to complete the study on a laptop or desktop in a single sitting. The median completion time was approximately 35 minutes.

\subsection{Visualization Literacy Evaluation}

\begin{table}[h]
\centering
\begin{tabular}{lcccc}
\toprule
 & \multicolumn{2}{c}{\textbf{Single-Turn}} & \multicolumn{2}{c}{\textbf{Multi-Turn}} \\
\cmidrule(lr){2-3}\cmidrule(lr){4-5}
\textbf{Item} & \textit{M} & \textit{SD} & \textit{M} & \textit{SD} \\
\midrule
Obstructive--Supportive & 0.92 & 1.63 & 0.96 & 1.63 \\
Complicated--Easy       & 0.84 & 1.72 & 0.65 & 1.86 \\
Inefficient--Efficient  & 0.86 & 1.80 & 1.00 & 1.76 \\
Confusing--Clear        & 1.15 & 1.70 & 0.60 & 1.79 \\
\midrule
\textbf{Mean}           & \textbf{0.94} & \textbf{0.13} & \textbf{0.80} & \textbf{0.19} \\
\bottomrule
\end{tabular}
\caption{UEQ-S pragmatic subscale item ratings across conditions (transformed to $-$3 to $+$3 scale), where M = Mean and SD = Standard Deviation. The mean row reflects the subscale mean and SD across the four pragmatic items.}
\label{tab:ueqs}
\end{table}

Table~\ref{tab:ueqs} shows the results for the pragmatic measures on the UEQ-S, which captures users' ratings for how usable and task-supportive our interface is. When we interpret the subscale means using the benchmark bands \cite{hinderks2017design}, both the single-turn ($M = 0.94$) and multi-turn ($M = 0.80$) conditions fall within the framework's established above-average band. Thus, users rated the pragmatic quality of both interfaces favorably compared to other interfaces.

\end{document}